\documentclass[preprint]{aastex}

\newcommand{\be}{\begin{equation}}
\newcommand{\ee}{\end{equation}}
\newcommand{\rmin}{ {R_{\rm min} }}  
 
\newcommand{\tauej}{ {\tau_{\rm ej} }} 
 
\newcommand{\awig}{ {\tilde a}} 
\newcommand{\munch}{ { \mu_{12}}} 

\begin{document}

\title{Dynamical Stability of Earth-Like Planetary Orbits in Binary Systems} 

\medskip
\author{Eva-Marie David$^1$, Elisa V. Quintana$^{2}$, Marco Fatuzzo$^1$, 
and Fred C. Adams$^{2,3,4}$}
\bigskip 
\affil{$^1$Physics Department, Xavier University, Cincinnati, OH 45207} 

\affil{$^2$Michigan Center for Theoretical Physics \\ 
Physics Department, University of Michigan, Ann Arbor, MI 48109}

\affil{$^3$Astronomy Department, University of Michigan, Ann Arbor, MI 48109}

\email{$^4$fca@umich.edu} 

\begin{abstract} 

This paper explores the stability of an Earth-like planet orbiting a
solar mass star in the presence of an outer-lying intermediate mass
companion. The overall goal is to estimate the fraction of binary
systems that allow Earth-like planets to remain stable over long time
scales. We numerically determine the planet's ejection time $\tauej$
over a range of companion masses ($M_C$ = 0.001 -- 0.5 $M_\odot$),
orbital eccentricities $\epsilon$, and semi-major axes $a$.  This
suite of $\sim40,000$ numerical experiments suggests that the most
important variables are the companion's mass $M_C$ and periastron
distance $\rmin$ = $a(1-\epsilon)$ to the primary star. At fixed
$M_C$, the ejection time is a steeply increasing function of $\rmin$
over the range of parameter space considered here (although the
ejection time has a distribution of values for a given $\rmin$). Most
of the integration times are limited to 10 Myr, but a small set of
integrations extend to 500 Myr. For each companion mass, we find
fitting formulae that approximate the mean ejection time as a function
of $\rmin$. These functions can then be extrapolated to longer time
scales.  By combining the numerically determined ejection times with
the observed distributions of orbital parameters for binary systems,
we estimate that (at least) 50 percent of binaries allow an Earth-like
planet to remain stable over the 4.6 Gyr age of our solar system.

\end{abstract}


\section{Introduction}          \label{sec:intro}

Most stars have companions. Recent discoveries of extrasolar planetary
systems have shown that Sun-like stars often have planetary mass
companions and that these extrasolar planets reside in a wide variety
of orbital configurations (Butler et al. 1999, Marcy et. al 2001;
Fischer et al. 2002; Tinney et al. 2002; see also the extrasolar
planet almanac\footnote{http://exoplanets.org/almanacframe.html}). In
addition, most solar type stars are known to reside in binary systems
(Abt 1983) and thus often have stellar mass companions; the
distributions of orbital parameters for these binaries are relatively
well known (e.g., Duquennoy \& Mayor 1991; hereafter DM91). If
terrestrial planets are also present in these solar systems, the
companions can affect their long term orbital stability. This paper
addresses this issue of planetary stability with the overall goal of
estimating the fraction of binary systems that allow an Earth-like
planet to remain stable over the current age of our solar system.

Although terrestrial planets have not been detected in extrasolar
systems (with main-sequence primaries) due to their small masses, they
are expected to form in planetary systems alongside their Jovian
counterparts (e.g., Ruden 1999; Lissauer 1993). 
The probable
existence of such planets motivates this present work and underlies
recent efforts to built the Terrestrial Planet Finder in the near
future.\footnote{http://planetquest.jpl.nasa.gov/TPF/tpf\_index.html}
To study the stability of Earth-like planets in these systems, we must
undertake a numerical investigation of the three-body problem.
Although the dynamics of systems containing three gravitationally
attracting bodies was considered over a century ago by Poincar\'e, an
exact analytic solution to the general problem is not possible. The
robust nature of these celestial systems is still being explored
numerically. Indeed, it is now understood that multi-body
gravitational systems generally exhibit chaotic behavior, thereby
making it impossible to draw specific universal conclusions about
their dynamics. A complete understanding of how a specific system
evolves can only be garnered through a thorough numerical
investigation. Furthermore, due to sensitive dependence on the initial
conditions, the results of any numerical study must be presented
statistically, in terms of the full distribution of outcomes resulting
from effectively equivalent starting conditions. In this context, we
consider equivalent starting conditions to be those with the same
masses, eccentricities, and semi-major axes for the orbits, but with
differing choices of initial phase angles (and other orbital elements
-- see Murray \& Dermott 2000). As we illustrate in greater detail
below, the ejection time displays a distribution of values for an
ensemble of effectively equivalent initial conditions.

A great deal of work, both analytical and numerical, has already been
done on stability (e.g., Szebehely 1980) and the development of chaos
in celestial mechanics (e.g., Lecar, Franklin \& Holman 2001). But
many of the recent investigations of planetary stability focus on
specific astronomical systems and are relatively narrow in scope
(e.g., Benest 1996; Wiegert \& Holman 1997; Laughlin \& Adams 1999,
2000; Rivera \& Lissauer 2000; Rivera \& Haghighipour 2002; Dvorak et
al. 2003).  More general investigations of planetary stability in
binary systems have been carried out (e.g., Rabl \& Dvorak 1988;
Holman \& Wiegert 1999, hereafter HW99), but previous studies have not
found the fraction of binaries that allow for stable Earth-like
planets. In this paper, we re-examine the three-body problem by
considering the stability of Earth-like planets in the presence of a
companion with mass in the range 0.001 $< M_C/(1 M_\odot) <$ 0.5. This
mass range includes companions as small as Jupiter and as large as K
stars. The portion of parameter space (the $a-\epsilon$ plane) is
chosen so we can combine the numerical stability results with the
observed distributions of binary orbital parameters to estimate the
fraction of systems that allow Earth-like planets to remain stable
over long time scales.

The first result of this investigation is a determination of the
ejection time for Earth-like planets over an extensive range of the
companion's initial orbital eccentricity $\epsilon$ and semi-major
axis $a$. In this context, the Earth-like planet is assumed to have
the mass of Earth and starts in a circular orbit with a radius of 1
AU. To a reasonable approximation, we find that the ejection time
$\tauej$ depends primarily on the periastron distance $\rmin = a
(1-\epsilon)$ of the companion (for a given companion mass $M_C$).
Over the sampled regime of parameter space, the dependence of the mean
ejection time $\langle \tauej \rangle (\rmin)$ is well characterized
by straight lines in a semi-log plot in the $\tauej-\rmin$ plane, even
though the ejection time displays a distribution of values for a given
value of periastron. The ejection time thus shows an exponential
dependence on $\rmin$ over a particular range of periastron values.
For higher mass companions, the plots also display an inner regime,
where $\rmin$ is small and the ejection times are consistently short
(a few hundred years -- much shorter than any astrophysical [or
geological] time scales of interest). Although this work focuses on
the stability of Earth-like planets, scaling laws allow the results to
be applied in a broader context.

The second result of this investigation addresses the possible
habitability of extrasolar terrestrial planets (e.g., Kasting,
Whitmire \& Reynolds 1993; Rampino \& Caldeira 1994).  Approximately
two-thirds of main-sequence stars (of solar type) are found in
multiple systems (Abt 1983), and the binary frequency is even higher
for pre-main-sequence stars (Ghez, Neugebauer, \& Matthews 1993).
These companions can disrupt the orbit of an Earth. Although this
issue has been explored by several groups (e.g., Gehman, Adams, \&
Laughlin 1996; Laughlin, Chambers, \& Fischer 2002; Menou \& Tabachnik
2002), our numerical simulations shed further light on the subject.
Our numerical results indicate that distant stellar companions
(specifically, those with sufficiently large values of $\rmin$) will
not disrupt the orbits of Earth-like planets. This work extends the
range of parameter space studied previously and provides estimates for
the fraction of binary systems that allow habitable planets.

This paper is organized as follows. The numerical methodology is
described in \S 2, along with considerations of both Hill stability
and scaling laws. The results from this suite of numerical experiments
are presented in \S 3, along with an analytic characterization of the
dependence of the mean ejection time $\tauej$ on the periastron value 
$\rmin$. In \S 4, we use the numerically determined ejection times in
conjunction with observed binary parameters to estimate the fraction
of binary systems that can contain habitable planets. The paper
concludes in \S 5 with a summary and discussion of these results,
including an application to the stability of Earth-like planets in
known extrasolar planetary systems.

\section{Methodology} 

Although gravitationally interacting, three-body systems have been
studied at length, their deceptively complicated nature has stymied
efforts toward a complete solution.  In this paper, we explore one
special case of the three-body problem (the results can be scaled to
other parameter choices -- see below). Specifically, we consider the
possible ejection of an Earth-like planet that starts in an initially
circular orbit with radius 1 AU (around a primary star with one solar
mass). Through long term dynamical interactions with an outer
companion, the orbital elements of the Earth-like planet evolve,
generally in chaotic fashion, until the planet is ejected from the
system.

In order to explore this stability issue on intermediate time scales,
we use two different types of numerical codes to perform a series of
three-body experiments. For the first method, we use a symplectic
mapping code (written for Laughlin \& Adams 1999, following the lead
of Wisdom \& Holman 1991).  Using a relatively small time step of 4
days, we can maintain high accuracy over the course of our 10 Myr
integrations. In the second method, Newton's equations of motion are
integrated directly using a Bulirsh-Stoer (B-S) scheme (Press et
al. 1992).  Although direct integration is computationally more
expensive, it is accurate and explicit. For the systems at hand, our
B-S scheme incurs errors in relative accuracy of order 1 part in
$10^{11}$ per total time step (where each time step in the three-body
problem is variable, but has a typical value of about 10 days).

The Earth's initial orbit is always set to be circular ($\epsilon_E$ =
0) with radius $r=a_E=1$ AU. The mass $M_C$, eccentricity $\epsilon$
and semi-major axis $a$ of the companion body are then specified for
each run. For the vast majority of our numerical experiments, both the
Earth and the companion are started in the same plane. (We briefly
explore the case of non-coplanar orbits in follow-up simulations.)
With given values for the masses, semi-major axes, and eccentricities,
the initial conditions for the simulations must also specify the
orbital phases.  We consider numerical experiments with the same
binary orbital parameters ($M_C, a, \epsilon$) and a random 
distribution for the remaining phase angles (see Murray \& Dermott
2000).

For each set of initial conditions, we integrate the system forward in
time. For the sake of definiteness, and in order to cover a large
range of parameter space, we use ten million years as the upper limit
for the integration time.  These experiments give us either a time
scale for instability or a lower limit of ten million years on the
possible ejection time.  The planet is considered to be ejected if any
of the following conditions are met: The energy of the planet becomes
positive; the eccentricity of the planet exceeds unity; the periastron
of the planet becomes smaller than the stellar radius (assumed to be 1
$R_\odot$) so the planet is accreted; or the semi-major axis of the
planet exceeds a maximum value (taken here to be 100 AU).

For numerical experiments with the same binary properties ($M_C, a,
\epsilon$), the ejection time $\tauej$ varies with the choices of
orbital phases. Because the systems are chaotic, this variation is not
smooth, i.e., small differences in the starting phase angles can lead
to large differences in the resulting ejection times.  An exploration
of parameter space shows that the ejection time displays a log-normal
distribution for an ensemble of different phases, i.e., different
realizations of the same underlying problem (see Figure 1).  We have
found the distribution of ejection times using both the symplectic and
the B-S code and find the same distribution. Figure 1 shows the
resulting distribution for the case with binary companion mass $M_C$ =
0.1 $M_\odot$, eccentricity $\epsilon$ = 0.5, and semi-major axis $a$
= 5 AU. Also shown is a normal (gaussian) distribution (in $\log
\tauej$) with the same mean and width. Figure 1 illustrates several
properties of this stability problem that guide the rest of this
investigation: First, the width of the distribution is substantial
($\sigma$ = 0.51 for the case shown here). Second, both numerical
methods give essentially the same result. Third, the distribution is
log-normal so that we average our ejection times in $\log \tauej$ for
the remainder of the paper.

In addition to the two main numerical codes, a limited number of our
results were verified and extended using the MERCURY symplectic
integration package (developed by Chambers 1999). First, twelve sets
of simulations were repeated for a Jupiter-mass companion, with each
set containing three runs at constant periastron distance $\rmin$
(with varying eccentricities and semi-major axes); these runs were
done to make sure that the different codes give the same results.
Next, taking advantage of the increased speed of the MERCURY code, we
used it to extend a limited number of runs out to integration times of
100 Myr.  A more detailed examination was then performed for a range
of systems that remained stable over a 100 Myr time scale by extending
the integration time to 500 Myr and sampling over a wider range of
Jupiter's initial eccentricity.

The numerical results obtained here can be compared with analytical
results obtained previously. As a reference point, we use standard
theory to specify the condition for Hill stability of the system
(following Gladman 1993). We first define the dimensionless quantities 
\be
\gamma_j = [1 - \epsilon_j^2]^{1/2} \, , \qquad 
\eta_j = {m_j \over m_1 + m_2} \, , \qquad 
\munch = {m_1 + m_2 \over (1 M_\odot)} \, , 
\label{eq:defs} 
\ee
where the subscripts $j$=1,2 refer to the Earth-like planet and the
companion, respectively. The condition for Hill stability can then be
written 
\be 
[ \eta_1 + \eta_2/a_2 ] 
[ \eta_1 \gamma_1 + \eta_2 \gamma_2 \sqrt{a_2} ]^2 > 1 + 
3^{4/3} \eta_1 \eta_2 \munch^{2/3} \, , 
\label{eq:stable} 
\ee
where $a_2$ is the semi-major axis of the companion, expressed in
units where the semi-major axis of Earth is unity. In the present
context, we start Earth with an initially circular orbit so that
$\epsilon_1 = 0$ and $\gamma_1 = 1$. In the limit $m_1 \ll m_2$, the
above expression simplifies to the form 
\be 
\gamma_2 = [1 - \epsilon_2^2]^{1/2} > 
\eta_2^{-3/2} - {1\over2} a_2 \eta_1 \eta_2^{-5/2} - 
\eta_1 a_2^{-1/2} \eta_2^{-1} \, . 
\label{eq:stable2} 
\ee
Saturating this bound, we thus obtain the eccentricity $\epsilon_2$
required for the system to become Hill unstable (for a companion 
with a given mass $\eta_2$ and semi-major axis $a_2$).

This criterion predicts that three-body systems will become Hill
unstable for modest values of the eccentricity. For example, with a
Jupiter-like companion, the Earth becomes Hill unstable for
$\epsilon_2 > 0.1$, i.e., a periastron distance of 4.7 AU. As many
previous authors have found for this regime of parameter space (e.g.,
Valsecchi et al. 1984; Milani and Nobili 1983; Gladman 1993), systems
that are unstable according to the criterion [\ref{eq:stable}] may
live for an extraordinarily long time (e.g., longer than the age of
the universe) before showing any signs of instability (e.g., see also
HW99; Levison, Lissauer, \& Duncan 1998; Duncan \& Lissauer 1998;
Wisdom \& Holman 1991; Laughlin \& Adams 1999). In this context, if
the system is unstable according to the Hill criterion [\ref{eq:stable}], 
then it has a chance to decay (by ejecting a planet) in the long
term. The time scale for decay can be described in terms of a
half-life (e.g., Adams \& Laughlin 2003); in a large sample of
equivalent solar systems, half of the systems will decay by ejecting a
planet in a well-defined time, but any particular system could decay
over a wide range of times. If the half-life (or expectation value of
the decay time\footnote{The expectation value is the mean decay time
averaged over the probability distribution. For an exponential decay
law, the expectation value $\tau$ is related to the
half-life $t_{1/2}$ via $t_{1/2}$ = $\tau\; \ln 2$.}) is
longer than the expected lifetime of the star (typically billions of
years) or the age of the universe (12--14 Gyr), then the system can be
considered as stable for evaluating the prospects for the
survivability of Earth-like planets. In this paper, we study systems
that are Hill unstable with half-lifes in the approximate range $10^2$
yr $< t_{1/2} <$ $10^9$ yr.

The results of this investigation can be scaled to other parameter
choices. Here, we concentrate on the case of an Earth-like planet and
fix its starting radius at $r=a=1$ AU. The equations of motion,
however, have a radial scale invariance. As illustrated by the Hill
stability criteria in equations [\ref{eq:defs} -- \ref{eq:stable2}],
all length scales in the problem can be scaled by the semi-major axis
$a_1$ of the Earth's orbit. If we rescale the problem by changing
$a_1$ by a factor $\cal F$, then the resulting dynamics are the same,
except that the time scales differ by a factor of ${\cal F}^{3/2}$.
Similarly, we perform simulations using a value of 1.0 $M_\odot$ for
the mass of the primary. If we rescale the problem for an arbitrary
mass $m \equiv M_\ast/(1.0 M_\odot)$, the calculated time scales
change by a factor of $m^{-1/2}$ and are applicable to companion
masses that are rescaled according to $M_C \to m M_C$.  The
mass of the `Earth' that is being modeled by the simulations also
changes such that $M_E \to m M_E$. Because $M_E \ll M_C, M_\ast$,
however, the Earth acts essentially like a test particle and its mass
is of little consequence.

\section{Results} 

Using the methodology outlined above, we have performed a large number
of simulations of the Sun-Earth-companion system using both the
symplectic and B-S numerical integration schemes. The long term
evolution of these numerical experiments follows the same general
trend. Over the first several thousand years, the companion drives
Earth into an orbit with ever higher eccentricity. Because the
companion is much more massive than Earth, its orbital elements change
far less than those of the planet. In addition, Earth's orbital
eccentricity does not show a slow and steady increase, but rather
shows a cyclic and often chaotic pattern (for examples of this type of
behavior, see Laughlin \& Adams 1999; Rivera \& Lissauer 2000; Lecar
et al. 2001).  As a result, Earth's eccentricity is best described as
a distribution of possible values. As the Earth continues to orbit its
star, it samples all of the eccentricity values in this distribution,
but the distribution itself evolves with time. Figure 2 illustrates
this behavior for one particular case near the center of our parameter
space, i.e., a binary with companion mass $M_C$ = 0.1 $M_\odot$,
eccentricity $\epsilon$ = 0.5, and semi-major axis $a$ = 5 AU. The
figure shows the distributions of eccentricity for Earth near the
beginning of the integration, at two intermediate times, and near the
ending of the integration (just before ejection of the Earth-like
planet). In general, the Earth samples the distribution more quickly
than the distribution changes; this result is consistent with the
Lyapunov time being correlated with (but much shorter than) the
ejection time (see Lecar, Franklin, \& Murison 1992).  Over longer
times, however, the distribution of eccentricities grows broader and
samples larger and larger values of $\epsilon$. When the distribution
becomes sufficiently wide, the Earth stands a fair chance of entering
into an orbit of extremely high eccentricity, which ultimately leads
to instability. The Earth then either plunges into the star or is
ejected from the solar system altogether.

Using the symplectic integration scheme, we have performed an ensemble
of approximately 35,000 simulations of the Sun-Earth-companion system.
The semi-major axis of the companion is chosen to vary in even
increments from 1 AU (where Earth is ejected almost immediately) out
to 15 AU (for the 1 $m_J$ companion), 20 AU (for the 10 $m_J$
companion), and 80 AU (for the 0.1 and 0.5 $M_\odot$ stellar mass
companions). For each choice of semi-major axis, the eccentricity is
chosen to sample a range of periastron values from $\rmin$ = 1 AU out
to values so large that the ejection time always exceeds the 10 Myr
range of our integrations. For each choice of companion mass,
eccentricity, and semi-major axis, the remaining phase angles are
chosen from a random distribution. This sampling of parameter space
fills a broad band in the $a-\epsilon$ plane, as shown by the gray
scale plots in Figures 3 -- 6. For the portion of the $a-\epsilon$
plane to the upper left of the chosen region, the companion crosses
the Earth orbit and would lead to rapid ejection. For the portion of
the $a-\epsilon$ plane to the lower right of the sampled region, the
ejection times are longer than 10 Myr.

The results obtained using the symplectic code are presented by
ascending order of companion mass in Figures 3 -- 6. The top panel of
each figure depicts a grayscale plot of ejection times in the
$a-\epsilon$ plane, where the darker shades correspond to longer
ejection times.  Contour lines of constant periastron $\rmin =
a(1-\epsilon)$ are included for reference. A natural dependency of the
ejection time on $\rmin$ is clearly illustrated by the grayscale
plots, with the largest gradients occurring perpendicular to contour
lines. To elucidate this trend, we plot the ejection time versus
$\rmin$ for each companion mass in the corresponding lower panel.
Although the ejection time depends most sensitively on the periastron
value, the relevant initial conditions that describe the binary
continue to be the eccentricity and semi-major axis. In the Figures,
the star symbols represent the average value of $\log \tauej$ (for
each given value of $\rmin$) and the vertical bars represent one
standard deviation about the mean.  The width of the distribution (at
constant periastron) arises from two sources: (A) For a given type of
binary (i.e., given $\epsilon$ and $a$), the ejection time has a
distribution of values as shown in Figure 1. (B) For binaries with
differing values of $(a,\epsilon)$, but constant periastron $\rmin =
a(1-\epsilon)$, the ejection time has additional variation (e.g., see
HW99). Although the ejection times vary by an order of magnitude at a
given value of $\rmin$, a clear functional dependence for the mean
value can nonetheless be extracted from the results. Over the sampled
range of periastron values $\rmin$, the mean ejection time follows a
straight line in the semi-log plots.  In the following section, we
will use this property to extrapolate our results out to longer
ejection time scales.

We note that the importance of the periastron distance has been found
in related dynamical investigations.  In a study of scattering of
Trans-Neptunian objects by the planet Neptune (Holman, Grav, \&
Gladman 2001), the borders of the region where the bodies become
chaotic are approximately described by lines of constant perihelia
(that study also discusses the departures from this trend).  In a
related work (Duncan, Levison, \& Budd 1995), objects with perihelia
less than about 35 AU were found to be unstable, apparently due to
cumulative effects of random forces from Neptune (exerted near
perihelion). Finally, the ejection time in systems of terrestrial
planets can be modeled with an equation similar to our equation
[\ref{eq:tj}] (see Chambers, Wetherill, \& Boss 1996).

In an alternate set of simulations, we have performed an ensemble of
approximately 8000 numerical integrations using the B-S code. The
results from the sympletic code suggested that the periastron is the
most important variable for determining the ejection time. For the B-S
integrations, we chose even increments in periastron $\rmin$ of the
companion. For each value of $\rmin$, we sample the eccentricity over
the full range $0 \le \epsilon \le 1$ and perform multiple
realizations of the experiments using different (random) choices for
the remaining phase angles. This coverage of parameter space is
complementary to that used for the symplectic code. The symplectic
code was used to sample as much of the $a-\epsilon$ plane as possible.
The B-S code was used to sample fewer values of $(a,\epsilon)$, and
fewer values of the periastron, but each point in the $a-\epsilon$
plane was studied with more realizations of the various phase angles
in the problem. The ejection time, as a function of periastron,
follows the same general trend for both ensembles of numerical
experiments, as shown in the bottom panels of Figures 3 -- 6 (where
results from the B-S code are plotted as triangular symbols). The main
difference between the two explorations of parameter space is that the
latter displays a somewhat wider distribution of ejection times for a
given periastron value.  However, the expectation values of the
distributions are in good agreement.

To package these results in a useful format, we fit the numerically
determined mean ejection times with an empirical relation of the form 
\be 
\tauej = \tauej_0 \exp\Bigl[ \alpha (p  - 1) \Bigr] \, , 
\label{eq:tj} 
\ee
where $\tauej_0$ is a fiducial time scale, $p$ is the dimensionless
periastron distance $p \equiv (\rmin$/1 AU), and $\alpha$ is the
dimensionless fitting parameter (the slope of the lines in Figures 3
-- 6).  The variables for the functional fits are presented in Table 1
below, for the four companion masses $M_C$ considered here. The Table 
also lists the range in $\rmin$ over which the fitting formulae have 
been numerically determined. For the stellar mass companions, the
ejection time is extremely short (hundreds of years) for a small range
of periastron near 1 AU; the fit to equation [\ref{eq:tj}] thus begins
at larger periastron values.

{}
\bigskip
\centerline{\bf Table 1: Parameters for ejection time scaling laws} 
\medskip 
 
\begin{center}
\begin{tabular}{cccc}
\hline 
\hline
mass $M_C$ ($M_\odot$)& $\tauej_0$ (yr) & $\alpha$ & periastron range \\ 
\hline 
0.001 & 1800 & 9.8 $\pm$ 3.0 & 1 AU $< \rmin <$ 1.8 AU \\ 
0.01 &  400 & 6.8 $\pm$ 1.1 & 1 AU $< \rmin <$ 2.1 AU \\ 
0.10 & 110 & 4.1 $\pm$ 0.67 & 1.6 AU $< \rmin <$ 3.5 AU \\ 
0.50 & 0.64 & 4.7 $\pm$ 0.32 & 2.4 AU $< \rmin <$ 4.4 AU \\  
\hline 
\hline 
\end{tabular}
\end{center}  

Although the ejection time depends most sensitively on the variable
$\rmin$, this time scale $\tauej$ is basically a function of
$a$ and $\epsilon$.  Close inspection of Figures 3 -- 6 shows that
nearby points in the $a-\epsilon$ plane can lead to rather different
ejection times. In addition, systems with the same values of $a$ and
$\epsilon$ can display differing ejection times for varying starting
phases of the orbiting bodies -- as illustrated in Figure 1. These
systems are thus chaotic, in the technical sense, and display
sensitive dependence on their initial conditions.  This variation
leads to a range of values for $\tauej$, as quantified by the error
bars in the lower panels of Figures 3 -- 6. For a given value of
$\rmin$, the ejection time $\tauej$ displays a full distribution of
values. As mentioned earlier, the width of this distribution arises
from two sources. First, the distribution of ejection times has an
intrinsic spread for a given pair $(a,\epsilon)$, as shown in Figure
1. Second, we are averaging over many pairs $(a,\epsilon)$ with the
same periastron; since the ejection time can depend on both $a$ and
$\epsilon$ (e.g., HW99), this averaging increases the width of the
distribution as shown by the error bars in the lower panels of Figures
3 -- 6.  The fitting formulae found above (see Table 1) provide an
estimate for the expectation value of this distribution as a function
of $\rmin$ (where the expectation value is the mean value, averaged
over the underlying probability distribution of ejection times).

The results shown in Figures 3 -- 6 are in basic agreement with
previous work.  In a study of Alpha Centauri, test particles were
found to be stable for initially circular orbits within about 3 AU of
either star (Wiegert \& Holman 1997), where the integration time
was 32,000 binary periods. In this binary, the mass ratio is close to
0.5, the periastron distance is 11.2 AU, and the semi-major axis is 23
AU.  Scaled to the units of this paper, the boundary of stability
corresponds to a time scale of $\tau \approx 4.8 \times 10^5$ yr for
$p \approx 3.7$. This point falls just above the best fit line shown
in the bottom panel of Figure 6, but well within the allowed range.
In a more general study, HW99 found the critical radius $a_{cr}$ for
which a planet will remain stable for $10^4$ binary periods. For a
given binary mass ratio, the critical radius is a function of
eccentricity (see Table 3 of HW99). When scaled to the
Sun-Earth-companion systems considered here (with one year planetary
orbits), the function $a_{cr}(\epsilon)$ can be converted to a
function $\tauej(p)$, where $\tauej$ corresponds to $10^4$ binary
orbits. The resulting function $\tauej(p)$ includes only one
eccentricity value for a given value of $p$, rather than an average
over a path in the $a-\epsilon$ plane as we use here.  Nonetheless,
the RMS value of the relative difference\footnote{This RMS difference
is calculated using equation (1) of HW99 and averaging over the range
of eccentricity $0 \le \epsilon \le 0.8$, i.e., the range listed in
their Table 3.} (in $\log \tau$) between the HW99 result and that
predicted by equation [\ref{eq:tj}] is only about 7\% for a companion
mass $M_C$ = 0.1 $M_\odot$ and 12\% for $M_C$ = 0.5 $M_\odot$.

The results from our lowest companion mass simulations were confirmed
for a limited number of cases using the MERCURY integration package
(Chambers 1999). For systems with a Jupiter-mass companion,
integrations were performed for periastron $\rmin$ ranging from 1.04
-- 2.60 AU.  For some of these simulations, the companion began in the
same orbital plane as the Sun-Earth system (as before). We also ran
an alternate series of simulations in which Jupiter began with an
inclination of 1.3$^{\circ}$ relative to the Sun-Earth orbital plane;
for comparison purposes, identical sets of simulations were also
performed treating the Earth as a mass-less test particle. In all of
the simulations, the other orbital elements were chosen at random.
Each system's evolution was then followed for 100 Myrs or until the
Earth was lost from the system.

For simulations with a starting inclination angle for Jupiter, the
results for periastron $\rmin$ = 1.04 AU show an interesting departure
from the co-planar simulations. These simulations (where Jupiter
starts with $i$ = 1.3$^{\circ}$) display a wider range of ejection
times, but the average value remains essentially the same. Although
the orbit of the companion brings it alarmingly close to Earth's
orbit, some of the simulations show relatively long ejection times. 
For example, in a simulation with Jupiter's $\rmin$ = 1.04 AU and
$\epsilon$=0.83, after 20,000 years, the (massive) Earth has been
perturbed into an orbit with $a$ = 9.2 AU, $\epsilon$ = 0.6, and an
inclination of $i$ = 139$^{\circ}$.  The Earth's eccentricity and
inclination continue to oscillate for several hundred thousand years
until the Earth is ejected from the system.

The MERCURY results verify that the range 1.6 AU $< \rmin <$ 1.8 AU
marks the transition between instability and stability. Most of the
systems with $\rmin <$ 1.7 AU lose the Earth in less than 10 Myr,
whereas systems with $\rmin >$ 1.9 AU remain stable for 100 Myr. The
stability of the high $\rmin$ systems was examined further through a
series of 500 Myr runs using $\rmin$ = 1.87 -- 2.01 AU and Jupiter's
eccentricity $\epsilon_J$ = 0.1 -- 0.8. Many of the systems remained
stable for the duration of the integration, although there were
several exceptions: With $\rmin$ = 1.87 and $\epsilon$=0.1, for
example, the Earth was ejected from the system in only 0.3 Myr.
However, the other systems with $\epsilon$ = 0.1 and $\rmin > 1.87$
remained stable for the full 500 Myr, as did the systems with 0.1 
$\le \epsilon \le$ 0.7.  On the other hand, half of the systems with
$\epsilon$ = 0.8 became unstable, with Earth being ejected from the
system or accreted by the star. Taken together, these results are
consistent with the ejection times predicted by the fit of equation
[\ref{eq:tj}], although the ejection time may be even longer (at large
values of periastron) than that predicted by the fitting function. A
larger ensemble of long term integrations is needed to clarify this
issue. One should also keep in mind that equation [\ref{eq:tj}]
represents the expectation value of the distribution; for a given
value of $\rmin$, the distribution has a width of nearly an order of
magnitude in ejection time $\tauej$.

\section{Fraction of binary systems that allow Earths} 

This set of numerical experiemnts can be used to address the question
of habitable planets. Although habitability includes many aspects, a
key requirement is for the planet to remain stable over long time
intervals.  The majority of stars reside in binary systems and these
companions can preclude the possibility of an Earth-like planet, i.e.,
a small planet in a 1 AU (nearly circular and stable) orbit about a
solar type star. The results of the previous section indicate that the
survival time of such an Earth increases with the periastron distance
$\rmin$.  In this section, we extrapolate our numerical results to
longer times and find an approximate lower bound on the fraction of
binary star systems that allow for habitable planets.

The numerical results of the previous sections are complete out to
ejection times of 10 million years and include limited results out to
500 million years. Although the time required for life to develop on a
planet is largely unknown, the most familiar habitable planet -- our
Earth -- has an age of 4.6 Gyr. Since we want to find a conservative
estimate for the fraction of binary systems that allow for
habitability, we assume that an Earth-like planet must remain stable
for 4.6 Gyr.

The ejection times for a given value of $\rmin$ span a wide range, as
shown in Figures 3 -- 6. Again adopting a conservative approach, we
assume that the ejection times take the shortest values within their
allowed range, and hence the ejection times are about 10 times shorter
than the numerically determined expectation values (as fit by equation
[\ref{eq:tj}]). Our estimate for the minimum periastron distance
required for survival over a time $\tau_{SS}$ thus takes the form 
\be
p > 1 + \alpha^{-1} \ln[10 \tau_{SS} /\tauej_0 ] \, , 
\label{eq:plimit} 
\ee 
where $p$ is the periastron $\rmin$ in units of AU. For the sake of
definiteness, we take $\tau_{SS}$ = 4.6 Gyr, the current age of the
solar system. In adopting the form [\ref{eq:plimit}], we are assuming
that equation [\ref{eq:tj}] continues to hold out to larger periastron
values and longer ejection times. Since our (limited) longer term
integrations indicate that Earth-like planets may survive even longer
than predicted by an extrapolation of equation [\ref{eq:tj}], the
limit implied by equation [\ref{eq:plimit}] represents a conservative
bound.  The parameters $\tauej_0$ and $\alpha$ are given in Table 1
for companions of various masses.

We note that this constraint can be generalized. If we consider an
Earth-like planet with circular orbit of radius $\awig$ = $a$/(1AU) in
orbit about a primary star of mass $m=M_\ast/(1 M_\odot)$, and we want
to enforce stability for a time $\tau$, the required constraint
becomes 
\be
p > \awig \Bigl\{ 1 + \alpha^{-1} 
\ln[(10 \tau/\tauej_0) m^{1/2} \awig^{-3/2} ] \Bigr\} \, . 
\label{eq:plimit2} 
\ee
The values of $\alpha$ and $\tauej_0$, as listed in Table 1, apply to 
rescaled companion masses $\widetilde{M}_C$ = $m M_C$ (where $M_C$ 
are the masses given in the table). 

Now we specialize back to our standard case of a 1.0 $M_\odot$ primary
and an initial Earth orbit of 1 AU. For companion masses $M_C$ = 0.1
and 0.5 $M_\odot$, the minimum value of the periastron distance
required for stability (according to equation [\ref{eq:plimit}]) is
about $p=6-7$, as derived from the range of allowed slopes listed in
Table 1. To be conservative, once again, we use the high end of this
estimated range.\footnote{We note that this limit is consistent with a
related result (HW99). Assuming that our solar system had a solar mass
companion, HW99 found the minimum semi-major axis required for
stability of the planets.  For $\epsilon$ = 0.4, they found that 
$a > 400 - 500$ AU.  Since the outermost planet is Neptune with
$a_{\rm Nep}$ = 30 AU, this result corresponds to a dimensionless
periastron in the range $p = 5.3 - 6.6$, consistent with the
constraint discussed here.} To find a lower limit on the fraction of
binaries that allow for habitable Earths, we thus need to find the
fraction of binary systems with dimensionless periastron $p > 7$.

The observed population of binaries has a well-defined period
distribution (DM91), which takes a log-normal form. Focusing on the
case of primary stars with masses of approximately 1.0 $M_\odot$, we
can convert the period distribution into the probability distribution
for semi-major axes $a$. The resulting normalized distribution thus 
takes the form 
\be
dP_a = f(\ln a) d \ln a \, = \, {1 \over \sqrt{2 \pi} \, \, \sigma} 
\, \exp \Bigl[ - {(x - x_0)^2 \over 2 \sigma^2} \Bigr] \, dx \, , 
\ee 
where the variable $x$ is the logarithm of the dimensionless
semi-major axis. The parameters $x_0$ and $\sigma$ are determined from
fits to the observed binary period distribution (as reported in DM91).
The dimensionless width $\sigma$ = 3.53. To convert the period
distribution to a distribution of semi-major axis, we must include
the ratio $\mu$ of the companion mass to the primary mass through the
relation $x_0$ = 3.44 $+ (1/3) \ln [1 + \mu]$. As a result, the 
distribution depends (weakly) on the value of the mass ratio $\mu$. 

The distribution of eccentricity is also well defined, but the
observed distribution takes a different form for close, intermediate,
and wider binaries (DM91). In the present context, we are interested
in binaries with periastron distances greater than about 7 AU, so the
semi-major axes are also greater than 7 AU and we can use the
eccentricity distribution for wider binaries. In this regime, the
eccentricity distribution has the simple form $d P_\epsilon = 2
\epsilon d\epsilon$ (DM91). Furthermore, for these wider binaries, the
eccentricity is independent of semi-major axis (DM91, Heacox 1998).

For a given mass ratio $\mu$, we define $F_\mu(p)$ to be the fraction
of binary systems with dimensionless periastron greater than $p$. To
evaluate $F_\mu(p)$, we must integrate over the portion of the
$a-\epsilon$ plane with periastron $a(1-\epsilon)>p$, where the
integrand is weighted by both the eccentricity distribution and the
distribution of semi-major axes (described above). The eccentricity 
integration can be done analytically and the remaining integral --
which defines the fraction $F_\mu(p)$ as a function of $p$ -- takes
the form 
\be
F_\mu (p) = {1 \over \sqrt{2 \pi} \, \, \sigma} 
\int_{\ln p}^\infty dx \bigl(1 - p {\rm e}^{-x} \bigr)^2 
\exp[-(x - x_0)^2/2 \sigma^2] \, . 
\label{eq:intfpmu} 
\ee
The fraction $F_\mu(p)$ is a slowly varying function of the mass ratio
$\mu$. To estimate the fraction $F(p)$ of all binary systems for which
Earth-like planets can remain stable, we must find the weighted average
of $F_\mu(p)$, i.e., 
\be
F(p) \equiv \int_0^1 w(\mu) F_\mu (p) d\mu \, , 
\label{eq:intfp} 
\ee 
where the distribution of mass ratios $d P_\mu = w(\mu) d\mu$. This
distribution has been observed and is presented in DM91 (see Table 7
and Figure 10). The peak of the distribution occurs near $\mu$ = 0.3
and the majority of systems have mass ratios in the range $0.1 \le \mu
\le 0.5$.  We use DM91 to specify the distribution $w(\mu)$ of mass 
ratios and evaluate the integral of equation [\ref{eq:intfp}]. The 
resulting function $F(p)$ is shown in Figure 7. 

Figure 7 also shows a fit to the numerically determined result. The
fitting function is chosen to have the simple form 
\be
{\widetilde F}(p) = F_1 \exp \bigl[ - (a \xi + b \xi^2) \bigr] \, ,
\label{eq:fitfp} 
\ee 
where $F_1$ = 0.711, $a$ = 0.101, $b$ = 0.0287, and $\xi = \ln [p]$.
The function [\ref{eq:fitfp}] provides a good approximation to the
numerically determined result, with an absolute error less than about 
0.011 and a relative error less than 4 percent. More exact fits are
not warranted, as the observed distributions of binary orbital
parameters are not known to this accuracy. This function can be used
to estimate the fraction of binary systems with periastron greater
than any specified value within the allowed range $1 < p < 10^5$.

As argued above, stability of Earth over 4.6 Gyr requires a stellar
companion to have $p > 7$ ($\rmin > 7$ AU), where this value has been
estimated from numerical experiments using companion masses in the
range $M_C = 0.1 - 0.5 M_\odot$ (which are typical values -- see
DM91).  The fraction of binary systems that meet this constraint and
allow habitable Earth-like planets is 0.5 (or 50 percent). This
estimate should be regarded as a lower bound on the fraction.
Additional binary systems could allow for habitable Earths if the
orbits can be inclined. Notice also that this estimate does not
include close binary systems, those with separations $a \ll 1$ AU.
Although sufficiently close binary systems could allow Earth to remain
stable (see, e.g., HW99), the companion will necessarily reside within
1 AU of the Earth and the companion's radiative flux could affect
considerations of habitability. For the wider binaries considered
here, the requirement of dynamical stability demands that $\rmin > 7$
AU so that the companion is always farther than about 6 AU from
Earth. The additional radiative flux of the secondary is always less
than about 3\% of that of the primary. For a more typical companion
mass, say $M_C = 0.4 M_\odot$, the flux from the secondary is less
than 0.2\% of the total.

The numerical results of the previous section show that
Sun-Earth-companion systems can remain stable for relatively long
times -- such as the current 4.6 Gyr age of our solar system -- even
though they do not meet the analytic criteria for stability (see
equations [\ref{eq:defs}--\ref{eq:stable2}]). To illustrate this
point, Figure 8 shows the allowed region of the $a-\epsilon$ plane for
Earth-like planets in binary systems. The solid curves delineate the
region of the plane that allows Earth to remain stable for 4.6 Gyr, as
estimated here, whereas the dashed curves delineate the (much smaller)
region of the plane for which the system is Hill stable. The allowed
region is that below each curve. Figure 8 emphasizes that the
requirement of system stability over geological, biological, and even
cosmological time scales (equations [\ref{eq:plimit},
\ref{eq:plimit2}]) can be less restrictive and more relevant than the
requirement of Hill stability (equation [\ref{eq:stable}]).

\section{Conclusion} 

This paper presents numerical simulations of Earth-like planets in
binary systems. These binaries have solar mass primaries and companion
masses in the range 0.001 $M_\odot$ (Jovian mass) to 0.5 $M_\odot$ (K
stars), although the results can be scaled to other choices.  The
first result of this work is an estimate of the ejection times for
Earth-like planets in these systems (Figs. 3 -- 6). For a given
companion mass, the ejection time depends most sensitively on the
periastron distance $\rmin$. Although the ejection time shows a wide
range for a given value of $\rmin$, the overall trend is well-defined
and the expectation value of the distribution can be described by an
exponential function of $\rmin$. We have fit our numerical results,
for each companion mass, to find the mean ejection time as a function
of periastron (see equation [\ref{eq:tj}] and Table 1).

The second result of this work is an estimate for the fraction of
binary systems with Sun-like primaries that allow an Earth-like planet
to remain stable for a specified time period. Our numerical
experiments suggest that the requirement for stability can be written
in terms of a minimum periastron distance for the binary orbit. The
resulting constraint is provided by equations [\ref{eq:plimit} --
\ref{eq:plimit2}] (where the fitting parameters $\alpha$ and
$\tauej_0$ have been calculated for companion masses $M_C$ = 0.001 --
0.5 $M_\odot$). For the observed distributions of binary orbital
parameters (DM91), the fraction of binaries that have periastron
distances greater than a given value is specified by equations
[\ref{eq:intfp} -- \ref{eq:fitfp}] and Figure 7. Taken together, these
results imply that at least 50 percent of the binary systems allow an
Earth-like planet to remain stable for 4.6 Gyr (the current age of the
solar system). Because the ejection time is a much more sensitive
function of periastron than the fraction of binary systems (with
periastron greater than a given value), this estimate is quite robust;
for example, if we were to adopt the overly conservative stability
requirement that $\rmin > 20$ AU, the fraction of viable binary
systems would still be 40 percent. We also note that the condition of
system stability over billions of years is much less restrictive than
the requirement of Hill stability (see Figure 8).

This type of stability analysis can be applied to the observed
planetary systems now being discovered in association with nearby
stars (e.g., Mayor \& Queloz 1995; Butler et al. 1999, Marcy et
al. 2001, Fischer et al. 2002).  For a subset of the known extrasolar
planetary systems, those with $a > 1.5$ AU, we have performed
additional sets of numerical simulations using the masses and orbital
properties of the observed giant planets as the companions. An
Earth-like planet, with the mass of Earth and an orbital radius of 1
AU, is assumed to reside in each system; we then study its
intermediate term prospects for stability. Figure 9 shows the result
-- the estimated ejection time for Earth-like planets plotted as a
function of the observed periastron distance of the giant planet. Each
system displays a range of ejection times for the hypothetical Earth
(the vertical bars in the Figure show the standard deviation of this
distribution). The systems with periastron greater than about 1.8 AU
allow an Earth-like planet to survive for more than 10
Myr. Extrapolating this result to the age of the galaxy, about 10 Gyr,
we estimate that systems with Jupiter mass companions and periastron
greater than 2.8 AU would remain stable; for companions with 10
Jupiter masses, the minimum periastron distance is about 4 AU.

The parameter space available to this class of systems is enormous and
additional numerical work should be carried out. This paper provides
an exploration of the $a-\epsilon$ plane using integration times out
to 10 Myr, but longer term simulations should be done for $t \gg 10$
Myr. In the regime studied here, the ejection time varies according to
the exponential law of equation [\ref{eq:tj}]. At sufficiently large
value of periastron, however, chaotic motion should no longer occur
and the system should become stable. In addition, this work focuses on
co-planar orbits, although longer ejection times can be realized for
varying orbital inclinations. This work also studies only single
planets, whereas multiple planets systems can also be considered.
Under favorable circumstances, multiple planets can protect each other
from ejection (M. Holman, private communication); if the orbital
precession induced by the other planets occurs on a shorter time scale
than that induced by the binary companion, the perturbations of the
companion can be washed out.  This work only considers the stability
of Earth-like planets with outer binaries, where the Earth lies within
the orbit of the secondary; the case of inner binaries, with Earth
orbiting the binary pair, should also be investigated. Finally, this
paper studies the stability of three-body systems after the Earth-like
planet has formed. The presence of a binary companion can affect the
planetary formation process, and an investigation of this issue is
underway (Quintana 2003).

\bigskip 
\bigskip 
\centerline{\bf Acknowledgements} 
 
This study began as an REU project for EMD at the University of
Michigan. We would like to thank Matt Holman and Greg Laughlin for
useful discussions and an anonymous referee for many useful comments
that improved the paper. EMD and MF are supported by the Hauck
Foundation through Xavier University. EVQ and FCA are supported by
NASA through a grant from the Origins of Solar Systems Program and by
the University of Michigan through the Michigan Center for Theoretical
Physics.

\newpage 

\newpage 
\begin{figure}
\figurenum{1}
{\epsscale{0.90} \plotone{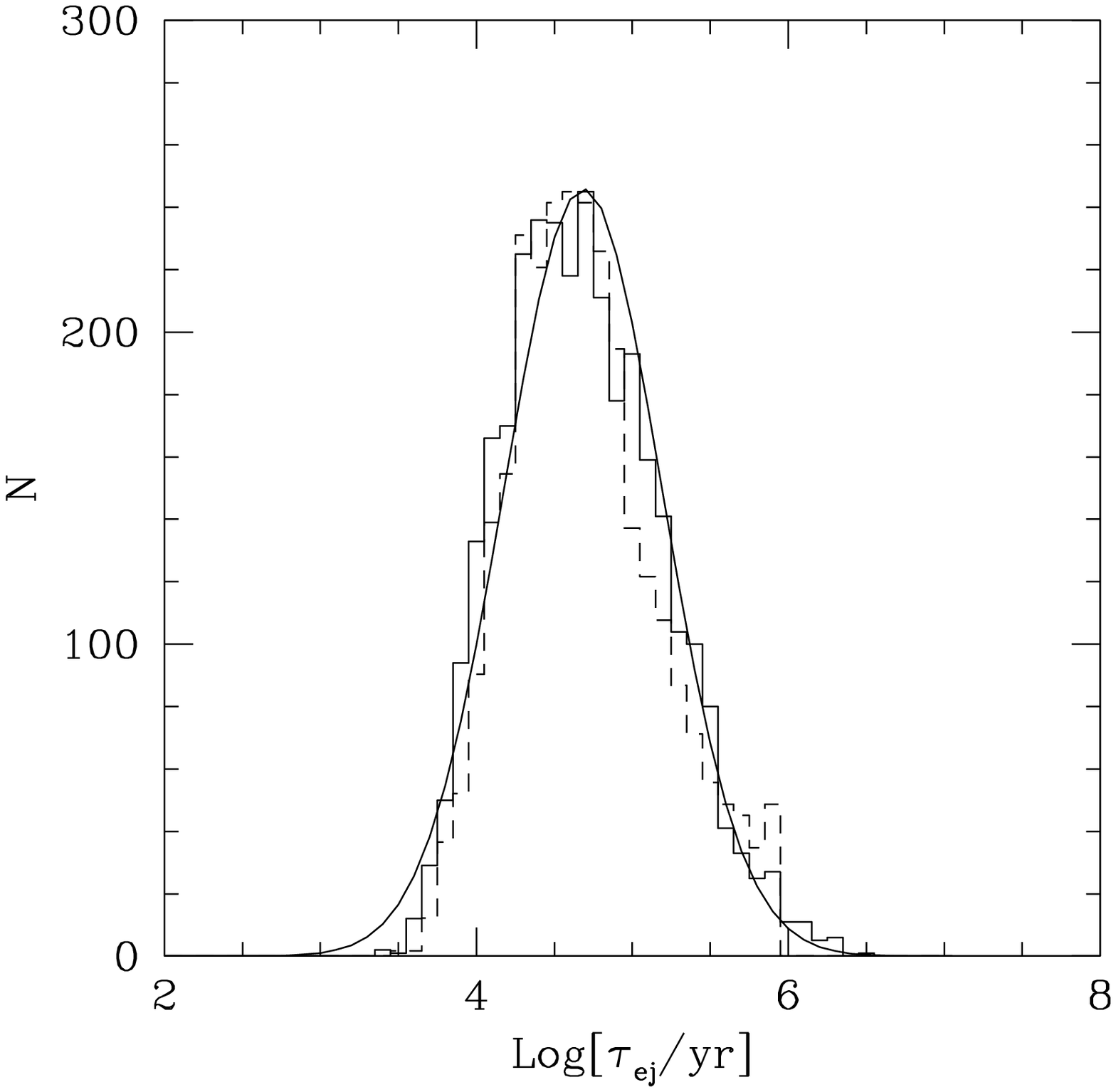} }
\figcaption{The distribution of ejection times for different
realizations of the same system. In this set of the experiments, 
the binary companion has mass $M_C$ = 0.1 $M_\odot$, eccentricity 
$\epsilon$ = 0.5, and semi-major axis $a$ = 5 AU. The solid histogram
shows the distribution of ejection times resulting from the B-S code
(for a random sampling of the starting phase angles). The dashed
histogram shows the corresponding distribution of ejection times
resulting from the symplectic code (again, for a random sampling of
phase angles). The smooth curve shows a log-normal distribution with
the same peak value and width as the computed distributions. Notice
that the distributions predicted by both numerical codes are similar
and that both have a log-normal form (with the same width and peak
location).}  
\end{figure}

\newpage 
\begin{figure}
\figurenum{2}
{\epsscale{0.90} \plotone{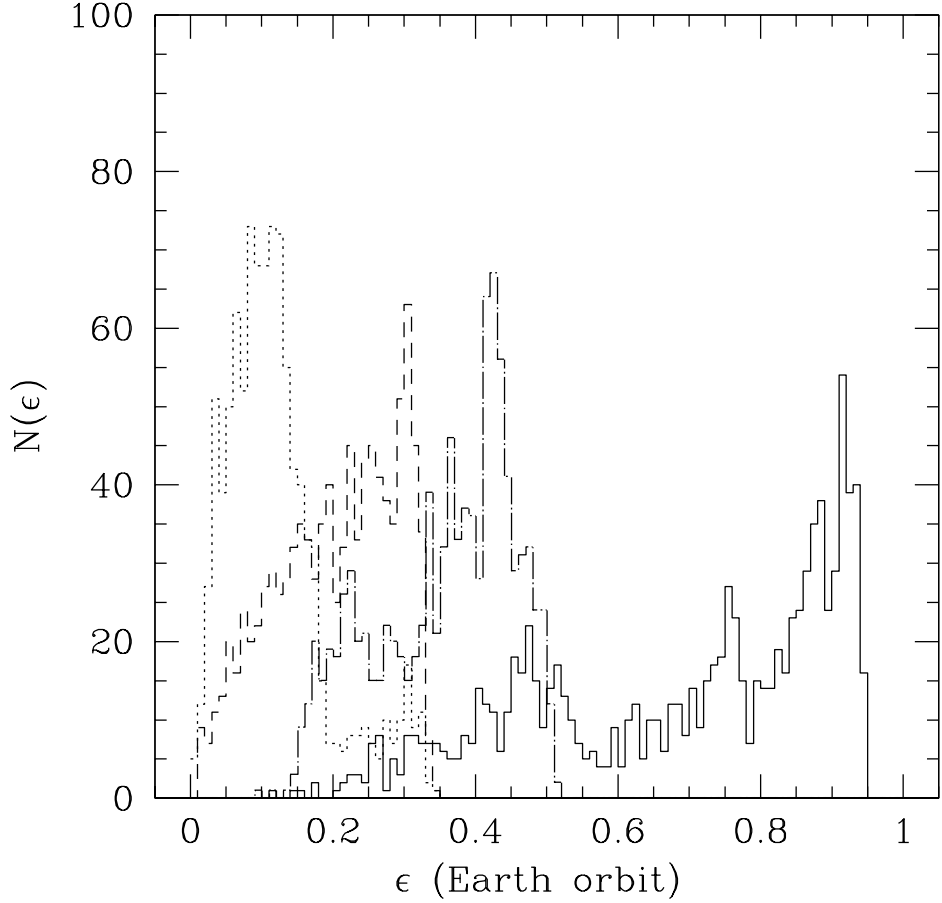} }
\figcaption{The distribution of eccentricity for the Earth-like planet
for different epochs of a single numerical integration. In this
experiment, the binary companion has mass $M_C$ = 0.1 $M_\odot$,
eccentricity $\epsilon$ = 0.5, and semi-major axis $a$ = 5 AU. The
dotted histogram on the left shows the distribution of eccentricity
over the first $10^4$ yr of the integration. The next three histograms
show the eccentricity distribution over $10^4$ yr intervals starting
at $5 \times 10^4$ yr (dashed curve), $8 \times 10^4$ yr (dot-dashed
curve), and the final interval ending at $\sim10^5$ yr (solid
curve). The eccentricity of the Earth-like planet thus displays a
distribution of values, and this distribution evolves toward higher
eccentricity values with time. } 
\end{figure}

\newpage 
\begin{figure}
\figurenum{3}
{\epsscale{0.5} \plotone{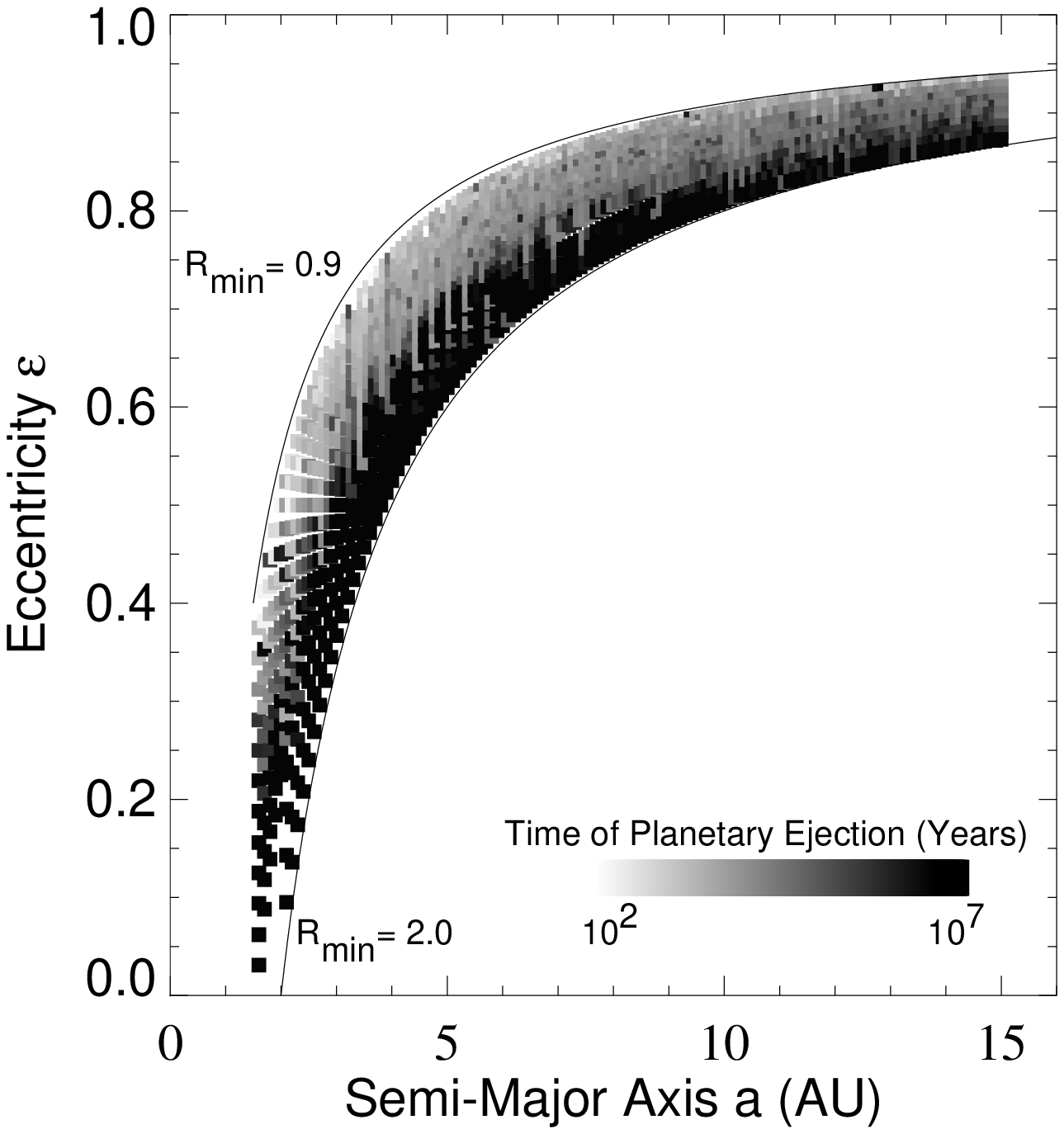} 
\vskip 0.12in
\centerline{\epsscale{0.5} \plotone{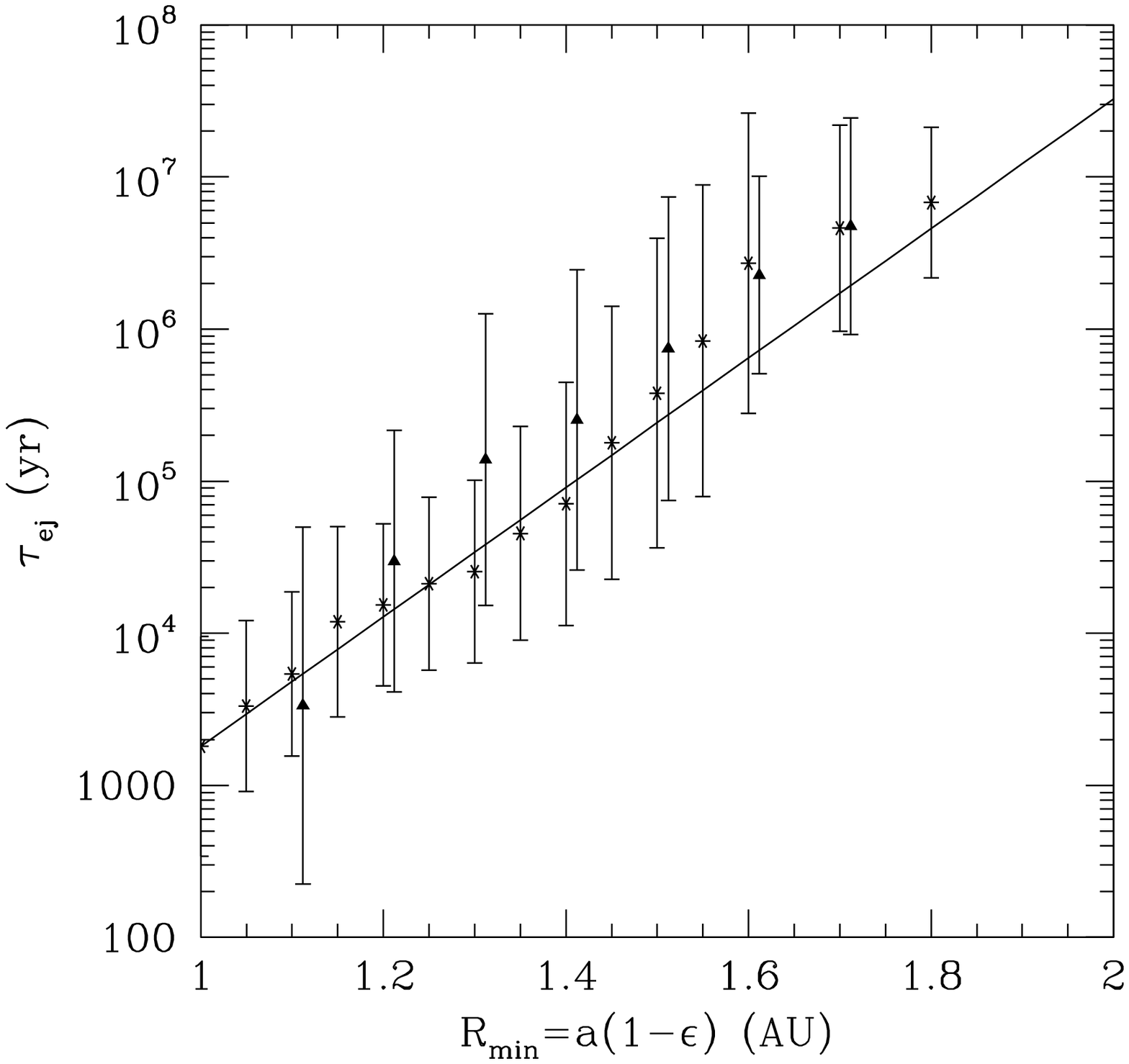} } }
\figcaption{Results of numerical simulations for $M_C = 0.001 M_\odot$. 
Top panel shows the gray scale plot of survival time as a function of
the location in the $a-\epsilon$ plane (using results from the
symplectic code). The lower panel shows the empirical form for the
survival time as a function of periastron distance $\rmin$. The
starred symbols show the results from the symplectic code, whereas the
filled triangles show the results from the B-S code (the triangles are
slightly offset for clarity).  }  
\end{figure}

\newpage 
\begin{figure}
\figurenum{4}
{\epsscale{0.5} \plotone{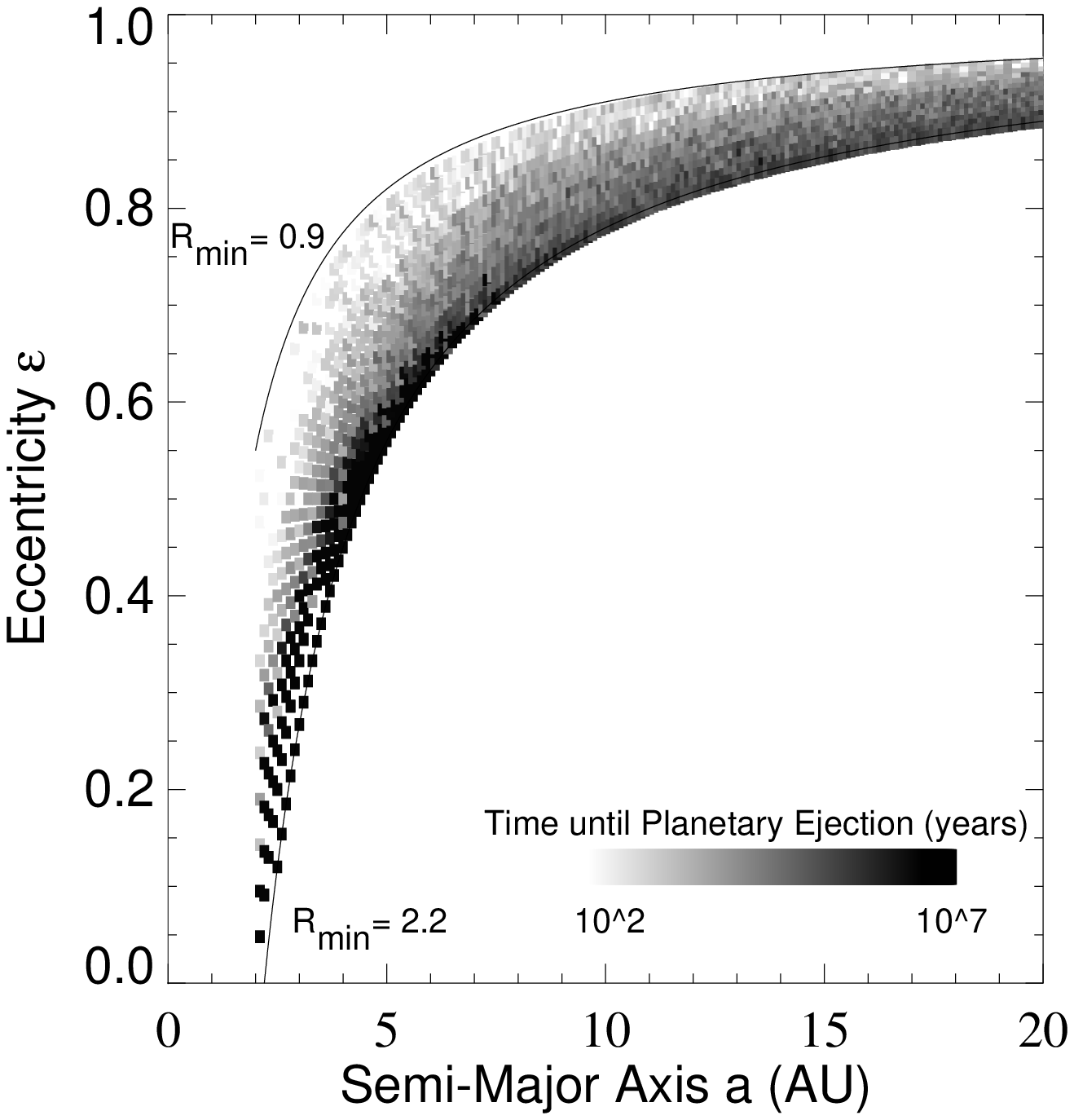} 
\vskip 0.12in
\centerline{\epsscale{0.5} \plotone{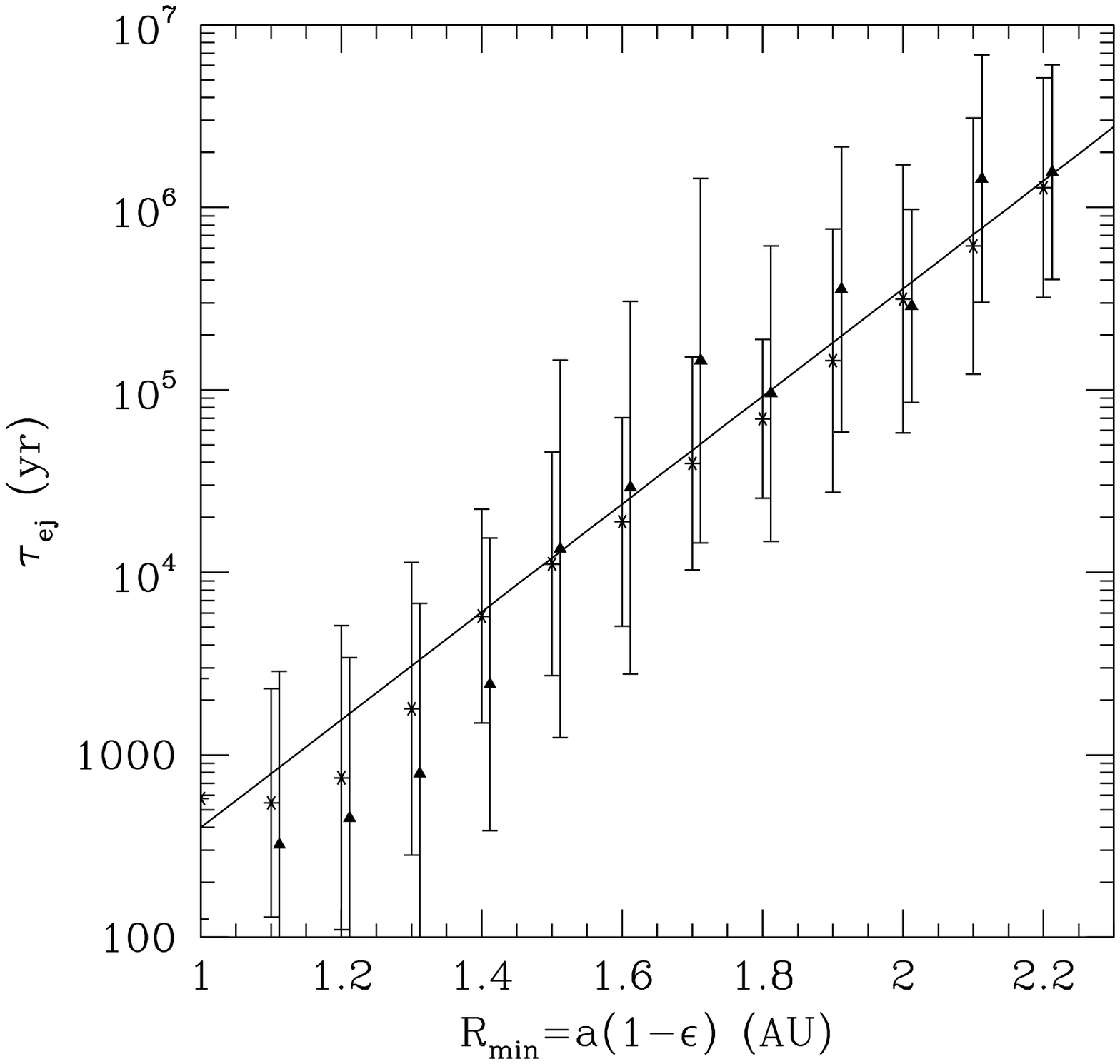} } }
\figcaption{Results of numerical simulations for $M_C = 0.01 M_\odot$.
Top panel shows the gray scale plot of survival time as a function of
the location in the $a-\epsilon$ plane (using results from the
symplectic code).  The lower panel shows the empirical form for the
survival time as a function of periastron distance $\rmin$.The starred
symbols show the results from the symplectic code, whereas the filled
triangles show the results from the B-S code (the triangles are
slightly offset for clarity). }  
\end{figure}

\newpage 
\begin{figure}
\figurenum{5}
{\epsscale{0.5} \plotone{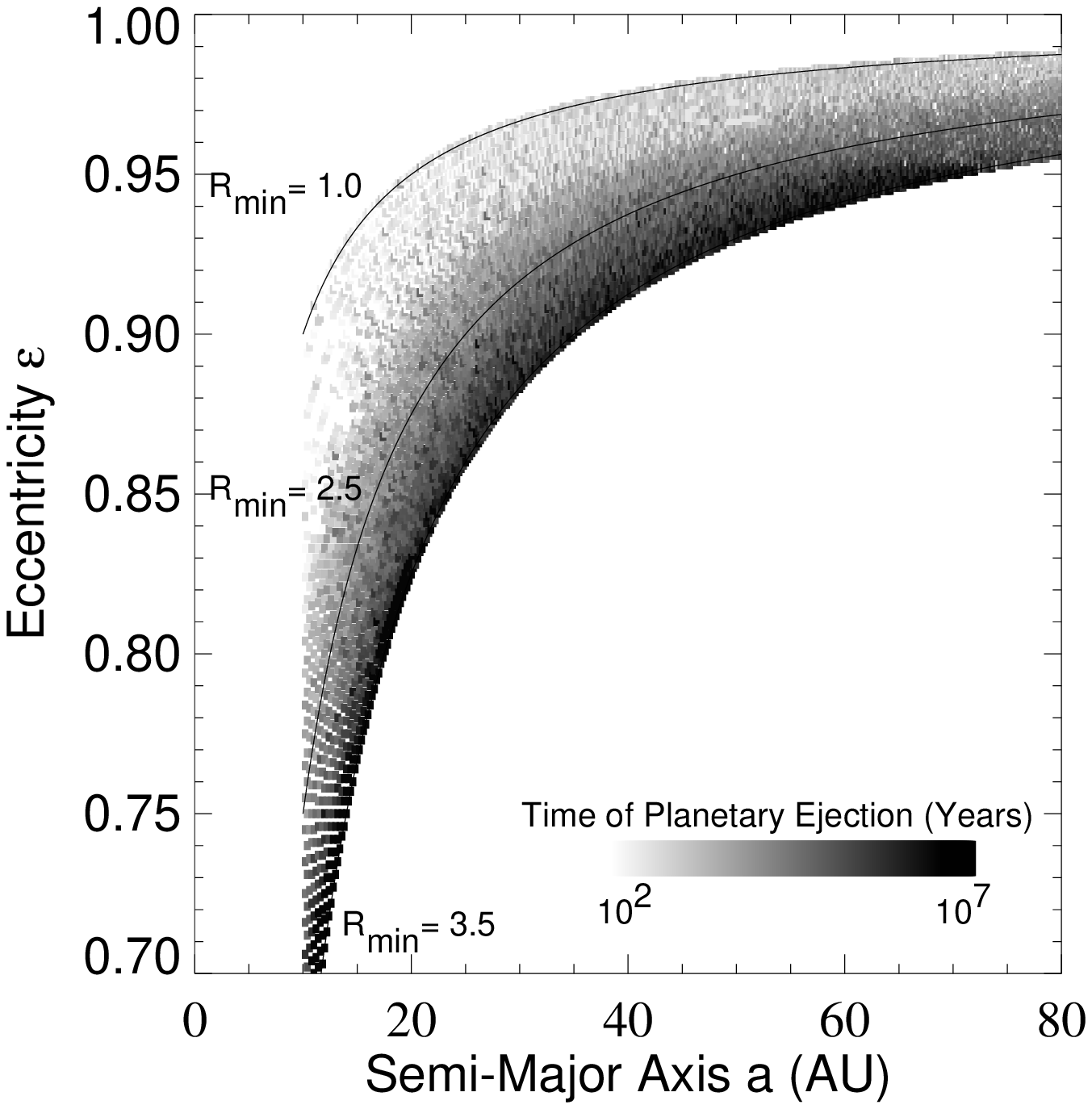} 
\vskip 0.12in
\centerline{\epsscale{0.5} \plotone{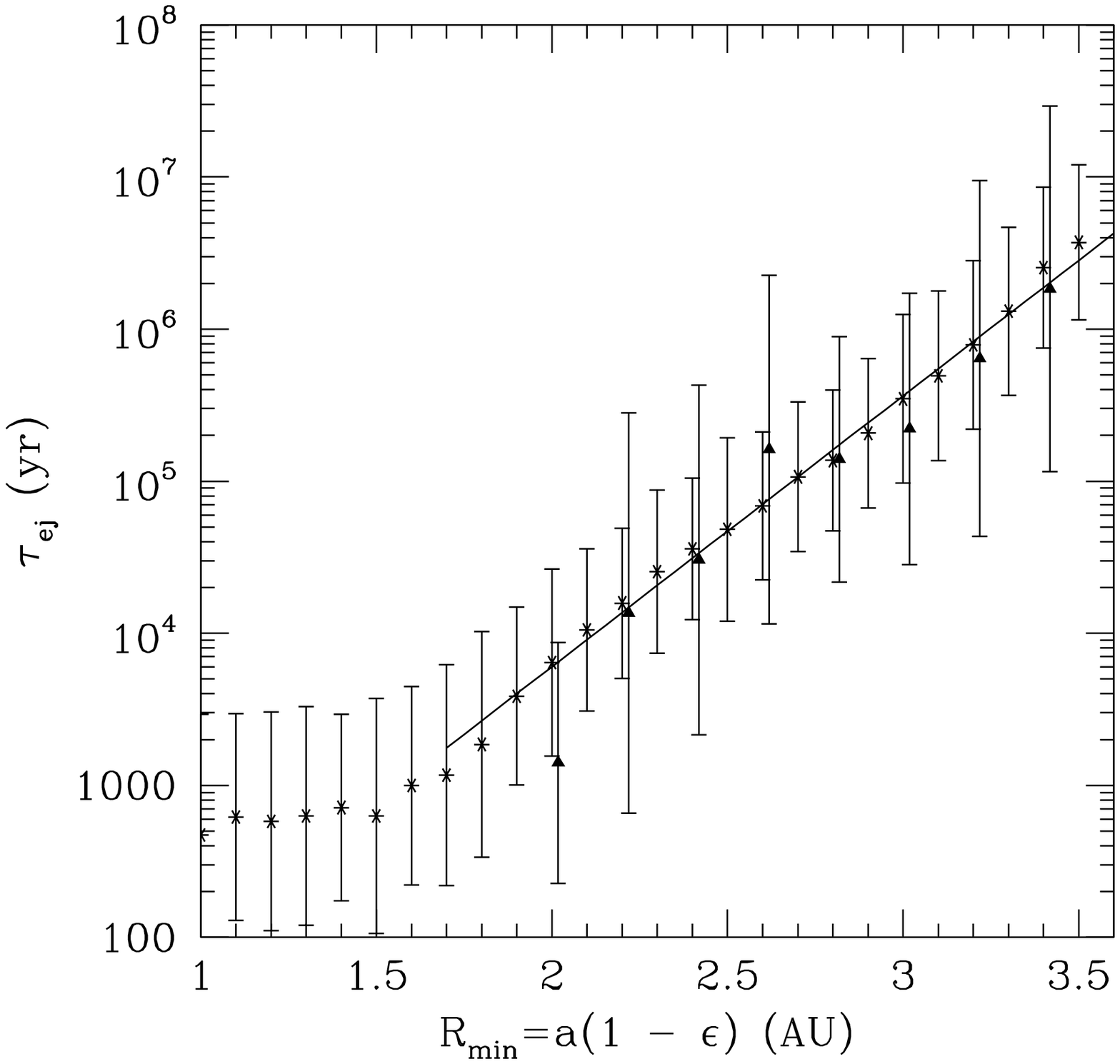} } }
\figcaption{Results of numerical simulations for $M_C = 0.1 M_\odot$.
Top panel shows the gray scale plot of survival time as a function of
the location in the $a-\epsilon$ plane (using results from the
symplectic code). The lower panel shows the empirical form for the
survival time as a function of periastron distance $\rmin$. Only the
top part of the plane ($\epsilon \ge 0.7$) is shown in the upper
panel, but the full range of $\epsilon$ was sampled to obtain the
slope of fitted line depicted in the lower panel.  The starred symbols
show the results from the symplectic code, whereas the filled
triangles show the results from the B-S code (the triangles are
slightly offset for clarity). }  
\end{figure}

\newpage 
\begin{figure}
\figurenum{6}
{\epsscale{0.5} \plotone{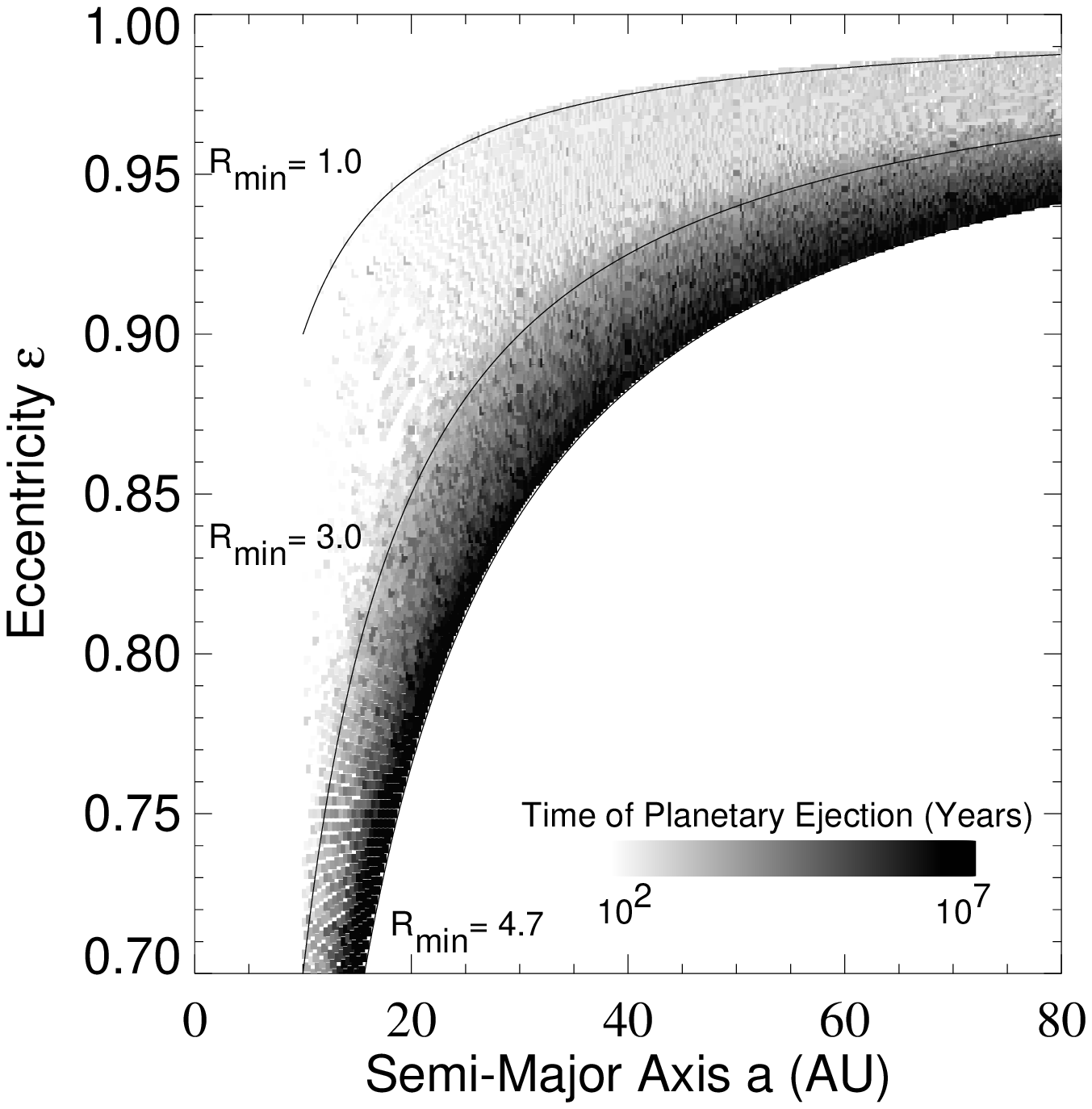} 
\vskip 0.12in
\centerline{\epsscale{0.5} \plotone{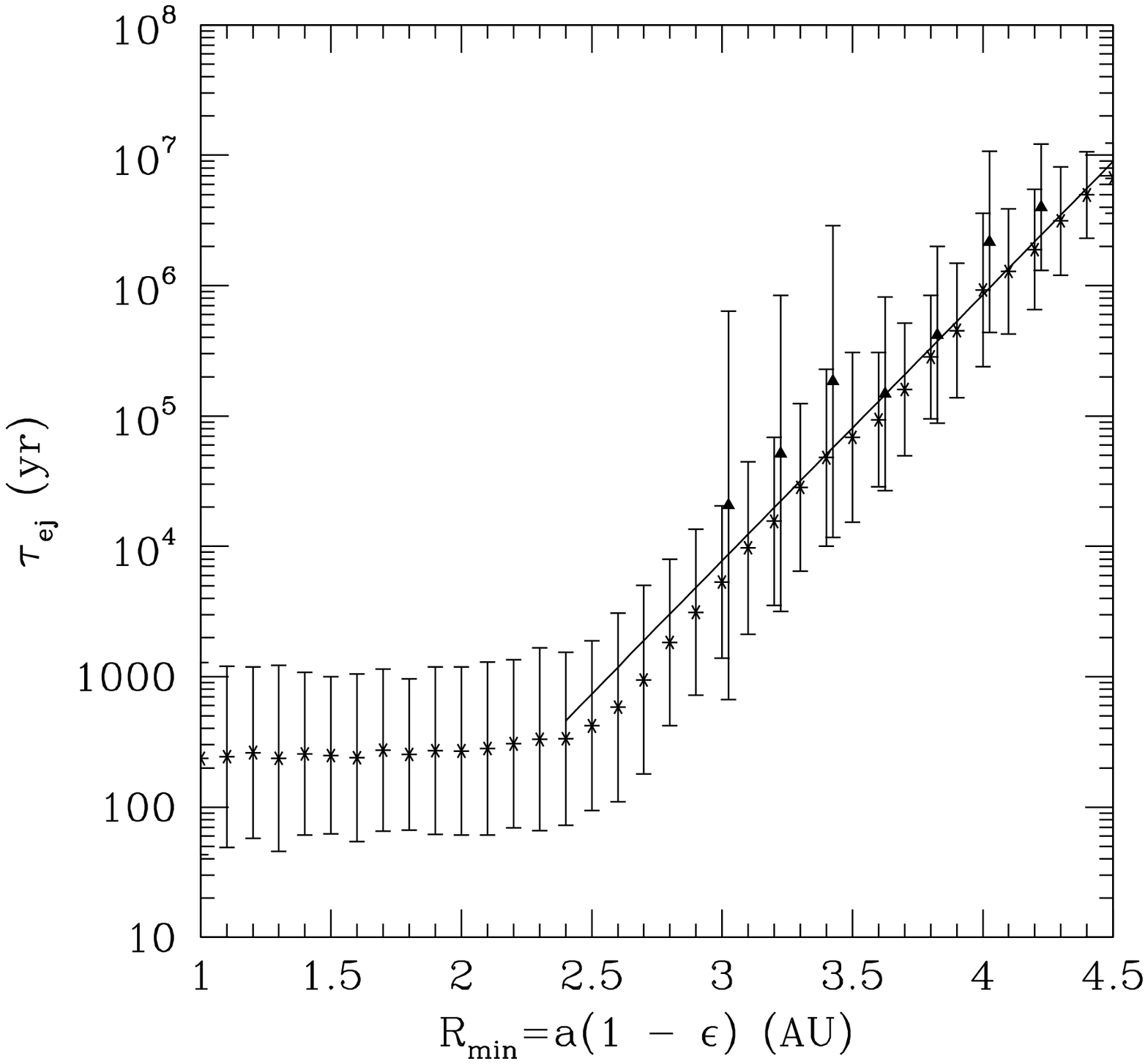} } }
\figcaption{Results of numerical simulations for $M_C = 0.5 M_\odot$.
Top panel shows the gray scale plot of survival time as a function of
the location in the $a-\epsilon$ plane (using results from the
symplectic code). The lower panel shows the empirical form for the
survival time as a function of periastron distance $\rmin$. Only the
top part of the plane ($\epsilon \ge 0.7$) is shown in the upper
panel, but the full range of $\epsilon$ was sampled to obtain the
slope of fitted line depicted in the lower panel.  The starred symbols
show the results from the symplectic code, whereas the filled
triangles show the results from the B-S code (the triangles are
slightly offset for clarity). }  
\end{figure}

\newpage 
\begin{figure} 
\figurenum{7} 
{\epsscale{0.90} \plotone{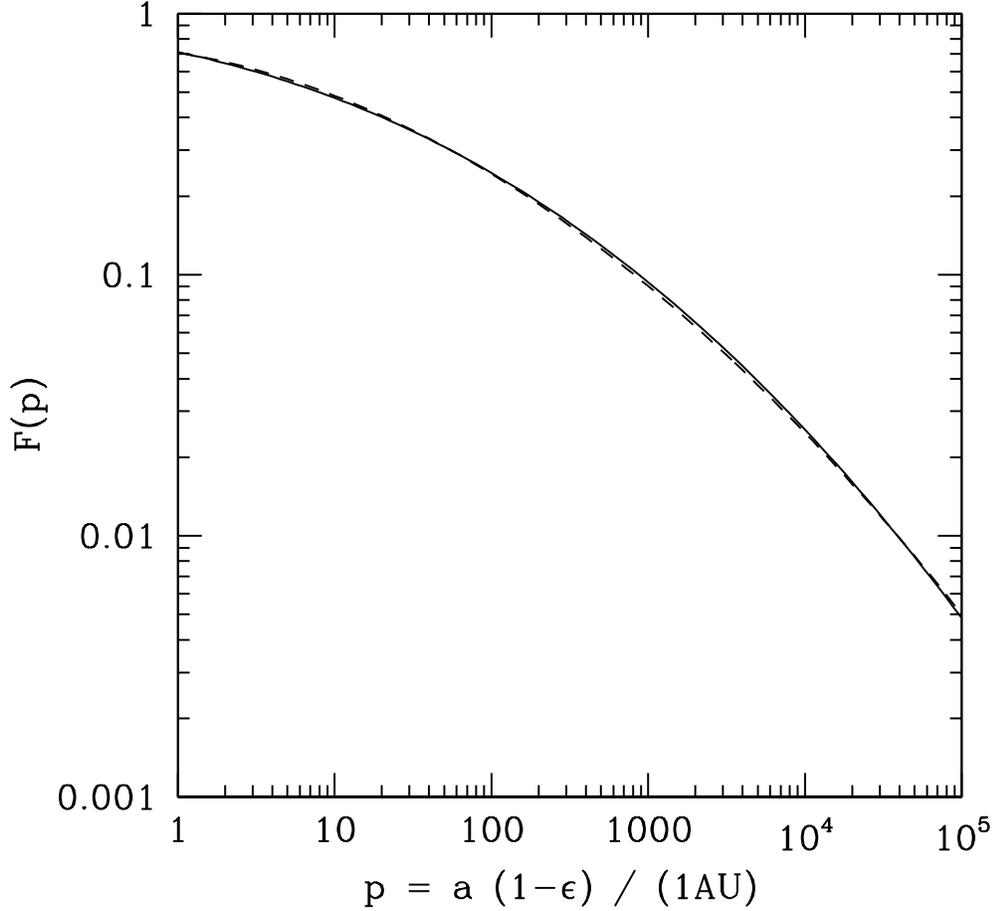} } 
\figcaption{Fraction of binary systems that have dimensionless
periastron distances greater than $p$, plotted as a function of $p$ =
$a(1-\epsilon)$/(1AU).  This fraction is determined from the observed
distributions of binary periods and orbital eccentricities (as 
reported in DM91). The dashed curve shows an analytic fit to the
numerical result (see equation [\ref{eq:fitfp}]). The numerical
simulations depicted in the previous figures indicate that the minimum
periastron required for Earth to survive 4.6 Gyr is about 7 AU. This
constraint, in conjunction with the distribution shown above, indicates 
that more than 50 percent of binary systems allow for habitable Earths.}  
\end{figure}

\newpage 
\begin{figure}
\figurenum{8}
{\epsscale{0.90} \plotone{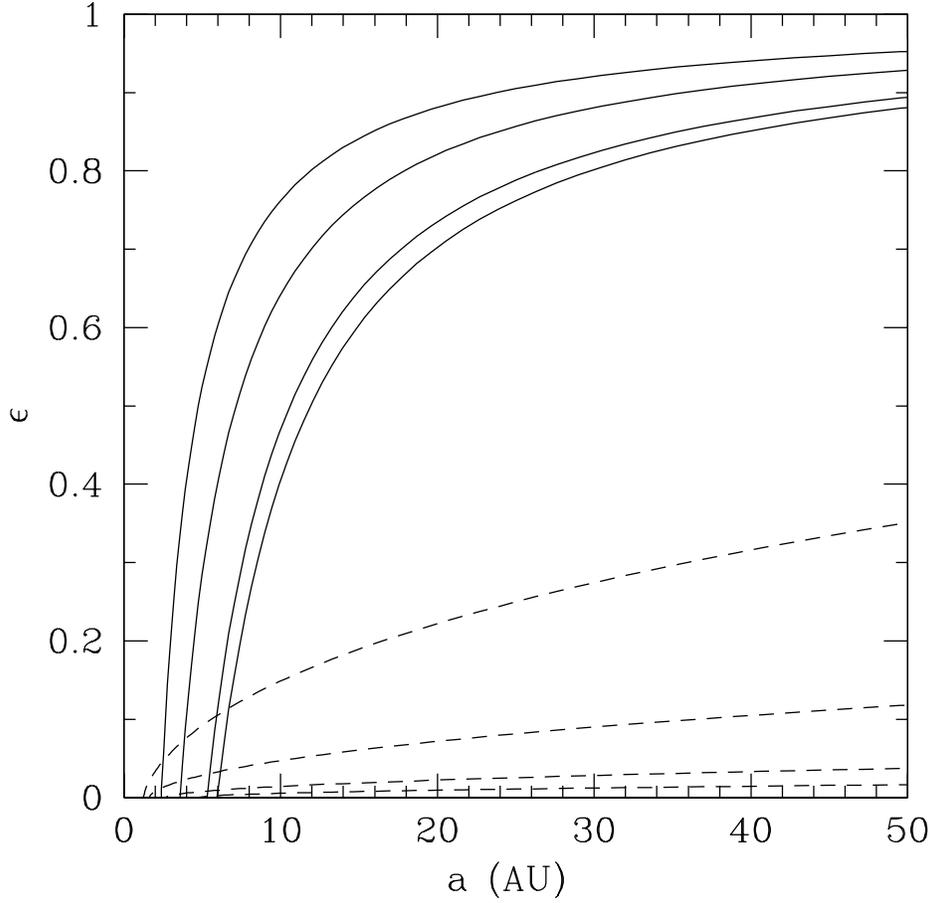} }
\figcaption{Comparison of analytic and numerical constraints on
survival time for Earth-like planets in binary systems with Sun-like
primaries.  The solid curves show the portion of the $a-\epsilon$
plane that allow for Earth-like planets to remain stable over the
current age of the solar system.  The allowed region of the plane
falls below the curves, which are displayed for companion masses $M_C$
= 0.001 $M_\odot$ (top), 0.01 $M_\odot$, 0.1 $M_\odot$, and 0.5
$M_\odot$ (bottom). The dashed curves show the analytic constraints
(eqs. [\ref{eq:defs} -- \ref{eq:stable2}]) that require the system to
be stable according to the Hill condition.  Again, the allowed region
falls below the curves, which are shown for companion masses $M_C$ =
0.001 $M_\odot$ (top), 0.01 $M_\odot$, 0.1 $M_\odot$, and 0.5
$M_\odot$ (bottom). Hill stability implies a much stronger constraint
than the requirement that Earth has a stable orbit for 4.6 Gyr.}
\end{figure}

\newpage 
\begin{figure}
\figurenum{9}
{\epsscale{0.90} \plotone{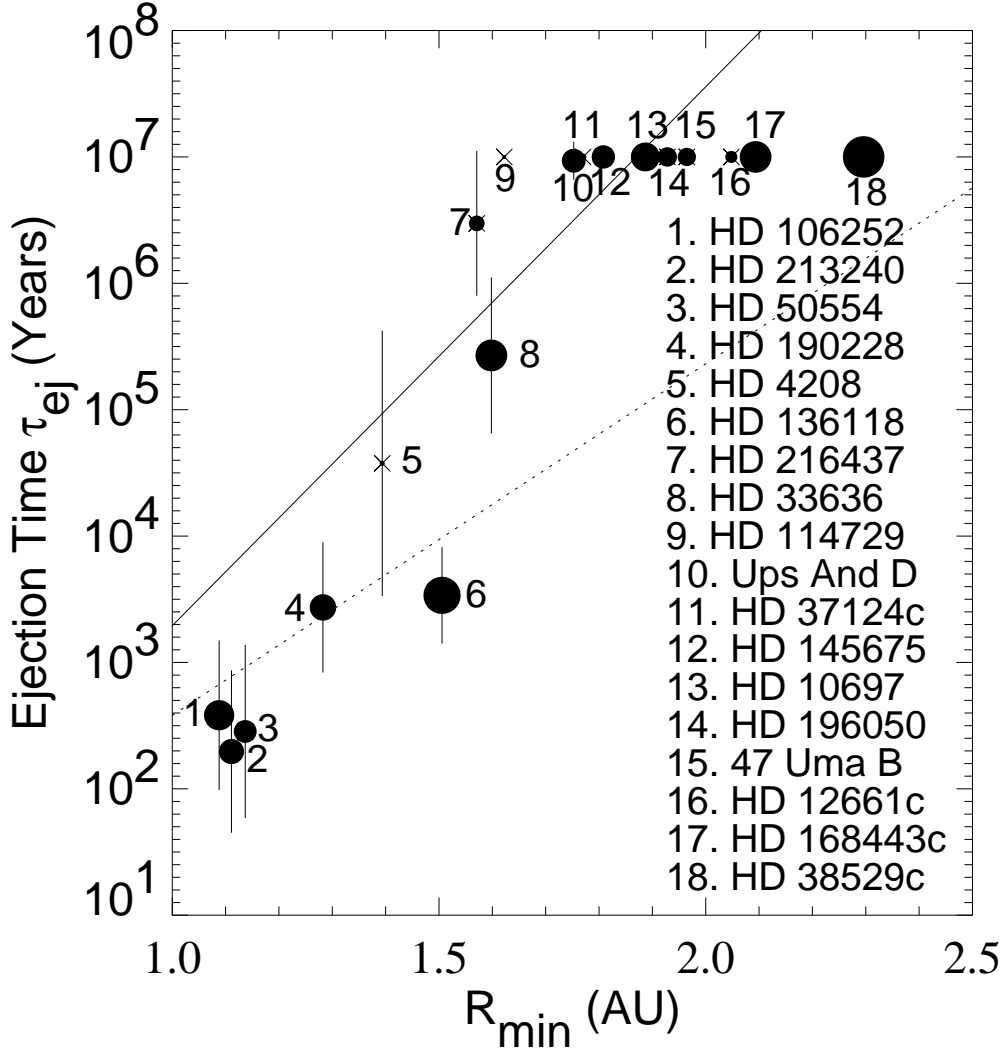} }
\figcaption{Results of numerical simulations for known extrasolar
planetary systems. The estimated survival time for Earth-like planets
is plotted as a function of the observed periastron distance of the
secondary (the extrasolar giant planet). The size of the plotting
symbols is proportional to the logarithm of the planet mass. The
vertical bars on each planetary symbol show the standard deviation 
for the distribution of ejection times for the hypothetical Earth in
the system. The solid curve shows the expectation value of the ejection 
time for a planetary companion with mass $M_C = 0.001 M_\odot$; the 
dashed curve shows the expectation value for $M_C = 0.01 M_\odot$.}  
\end{figure}

\end{document}